\documentclass[aps,reprint,amsmath,longbibliography]{revtex4-1} 

\bibpunct{}{}{,}{s}{}{}

\usepackage{dcolumn} 
\usepackage{graphicx} 

\usepackage{color}					

\begin{document}

\title{Linking Behavior in the PER Coauthorship Network \title[Linking Behavior in PER]{Linking Behavior in the PER Coauthorship Network}}
\date{\today}

\author{Katharine A. Anderson}
\email{andersok@andrew.cmu.edu}
\affiliation{Tepper School of Business, Carnegie Mellon University, Pittsburgh, PA 15213}
\author{Matthew Crespi}
\affiliation{Heinz School, Carnegie Mellon University, Pittsburgh, PA 15213}
\author{Eleanor C. Sayre}
\affiliation{Department of Physics, Kansas State University, Manhattan, Kansas 66506}

\begin{abstract}
There is considerable long-term interest in understanding the dynamics
of collaboration networks, and how these networks form and evolve
over time. Most of the work done on the dynamics of social networks
focuses on well-established communities. Work examining emerging social
networks is rarer, simply because data is difficult to obtain in real
time. In this paper, we use thirty years of data from an emerging
scientific community to look at that crucial early stage in the development
of a social network. We show that when the field was very young, islands
of individual researchers labored in relative isolation, and the coauthorship
network was disconnected. Thirty years later, rather than a cluster
of individuals, we find a true collaborative community, bound together
by a robust collaboration network. However, this change did not take
place gradually\textemdash the network remained a loose assortment
of isolated individuals until the mid-2000s, when those smaller parts
suddenly knit themselves together into a single whole. In the rest
of this paper, we consider the role of three factors in these observed
structural changes: growth, changes in social norms, and the introduction
of institutions such as field-specific conferences and journals. We
have data from the very earliest years of the field, a period which
includes the introduction of two different institutions: the first
field-specific conference, and the first field-specific journals.
We also identify two relevant behavioral shifts: a discrete increase
in coauthorship coincident with the first conference, and a shift
among established authors away from collaborating with outsiders,
towards collaborating with each other. The interaction of these factors
gives us insight into the formation of collaboration networks more
broadly.
\end{abstract}


\maketitle

\section{Introduction}

There is considerable interest in understanding the dynamics of coauthorship
networks---in particular, how changes in the culture and institutions
of a field affect the nature of its research community. This is important,
because there is evidence suggesting that the structure of knowledge
mirrors the social structure of the community producing that knowledge\cite{Traweek1988}.
Moreover, coauthorship connections are one part of a collaborative
system which facilitates (or impedes) the spread of information. Thus,
the progress of a field of research will be shaped by the patterns
of collaboration within it. Indeed, administrative and funding agencies
have spent considerable money and effort attempting to change existing
patterns of collaboration to improve researcher productivity and participation in science (e.g. the National Science Foundation programs to build community).

Coauthorship networks as a static entity have been studied extensively\cite{Grossman1995,Newman2001a,newman2001b,Newman2001PNAS,newman2004coauthorship}.
The dynamics of coauthorship networks have also been explored in a
range of academic fields, including mathematics (1940-1999 \cite{grossman2002evolution}),
sociology (1969-1999 \cite{Moody2004}), biotech (1988-1999 \cite{powell2005network}),
economics (1970-2000 \cite{Goyal2006}), network science (1998-2006
\cite{lee2008evolution}), and astrophysics (1998-1999, 2001-2006
\cite{heidler2011cognitive}). Unfortunately, due to data constraints,
these studies have largely focused on the dynamics of relatively established
academic communities. In this paper, we use thirty years of bibliometric
data from physics education research (PER) to look at how the structure of the coauthorship network
evolves during the crucial early stages in the development of an academic
community. Since 1981, the field of PER has grown dramatically, from
a handful of researchers, to hundreds of authors publishing over 150
articles per year. During this same period, the collaborative community
also changed, evolving from islands of individual researchers laboring
in relative isolation, into a true collaborative community, bound
together by a robust collaboration network. However, that change did
not take place gradually: it happened suddenly and dramatically in
the mid-2000s.

This prompts a question: what is the source of the changes we see
in this emerging community? During this period, two different institutions
are introduced: the first field-specific conference, and the first
field-specific journals. We also identify two relevant behavioral
shifts: a discrete increase in coauthorship coincident with the first
conference, and a shift among established authors away from collaborating
with outsiders, towards collaborating with each other. The interaction
of these factors gives us insight into the formation of collaboration
networks more broadly.

\section{Network Terminology}

For those unfamiliar with network terminology, it is valuable to define a variety of terms that we will use below. A network, $g$, consists of nodes (represented by circles) and links (represented by lines). Nodes are generally agents of some kind; in this case, they are authors in the field of PER. A link between nodes A and B indicates a relationship between the two agents; in this case, authors $i$ and $j$ are connected if they have coauthored a paper together. Each link has a weight, $w_{ij}$, representing the strength of the relationship.  In this case, we will weight links by the number of papers two authors have written together. The network at large consists of a number of \emph{connected components:} sets of nodes that can all be accessed by traveling across links in the network. 
The largest of the components is called the \emph{largest connected component} (LCC)
The \emph{degree }of node $i$, $d_{i}$, is the number of direct connections she has. In the case of a collaboration network, an author's degree is the number of coauthors she has. A node's\emph{ centrality} represents how important the node is in the collaboration network. There are many types of network centrality, which are interpreted differently. \emph{Degree centrality} is the simplest, representing the node's degree in the network, normalized by the maximum possible degree: $\frac{d_{i}}{\left(N-1\right)}$, where $N$ is the number of nodes in the network. \emph{Eigenvector centrality,} on the other hand, reflects the fact that nodes connected to important nodes are likely more important themselves. It is called eigenvector centrality because if we represent the network as a matrix, $A$, where $A_{ij}=1$ if $i$ and $j$ are connected, and $0$ otherwise, 
then the eigenvector centrality of node $i$ is the $i$th entry of the normalized eigenvector associated with the largest eigenvalue of $A$.
Eigenvector centrality may be different than degree centrality because an author connected to a few giants in the field will be more important than an author connected to a large number of unknown authors.  

\section{Data selection}

In this paper, we look at PER publications written between 1981 and 2010. This time period is particularly valuable, because it encompasses most of the early history of the field, including the introduction of several milestone institutions. It is also a period of dramatic growth and change in the community, making it an ideal window into the early life of the field. Arguably, PER has roots in the broader science education community, with intellectual parents in Dewey\cite{dewey1910science} and Arons \cite{Arons1976b,Arons1976}. However, in the US, PER has only been housed in departments of physics since the mid-1970s, and the research community took on an identity of its own substantially after 1980. Thus, starting our data collection in 1981 reasonably captures the development of the field.

Our data come from three journals: The American Journal of Physics (AJP), the Physics Education Research Conference Proceedings (PERC Proceedings), and Physical Review Special Topics---Physics Education Research (PhysRev-PER). Our data collection ended before the journal changed to its present name, Physical Review---Physics Education Research. These are the three most-common peer-reviewed outlets for PER in the US. While PER authors publish in other journals, and PER community members commonly read and cite articles from other sources, each of those other sources publishes fewer PER articles per year and is read by a much broader audience than these three. It's difficult to estimate how many other papers are written or read by PER community members because the tail on the publication venue distribution is very long. However, papers from these three journals comprise about half of all papers on PERticles, a community-supported reference aggregation group, aimed at recent papers of interest to PER readers. The full PERticles database is less relevant to our needs, because 1) it is not a complete listing of PER articles and 2) it predominately chronicles recent papers. Given that our interest is in early-stage development, we have chosen to use the full records of the top three journals instead.

For each journal, we use the bibliographic information for all relevant
articles. In the PERC Proceedings and PhysRev-PER, we assume that all content is PER-related.
In AJP, where much of the content is of more general physics interest,
articles are hand-coded by a member of the field (Steve Kanim)
to identify PER papers. This gives us a data set of 1114 PER papers:
276 in AJP, 481 in PERC Proceedings, and 226 in PhysRev-PER.  

Author names were reduced to first, middle initial, last and then
hand-disambiguated by a member of the field (Eleanor C Sayre). Authors
who changed their names during the relevant time period were listed
under their most recent name; no two authors in this time period had the same names as each other. 

One advantage of using early-stage data is that this period includes
several important field-specific milestones. Figure \ref{Number of papers Number of authors}
shows a timeline, including the PER workshop in 1997 (a precursor to the Physics Education Research Conference (PERC) introduced in 1998), the PERC Proceedings in 2001, and PhysRev-PER in 2005. This period also includes the appearance of several PER-specific
graduate programs, and the growth of National Science Foundation funding for PER.

\section{Network Size}

\begin{figure}
\includegraphics[width=\columnwidth]{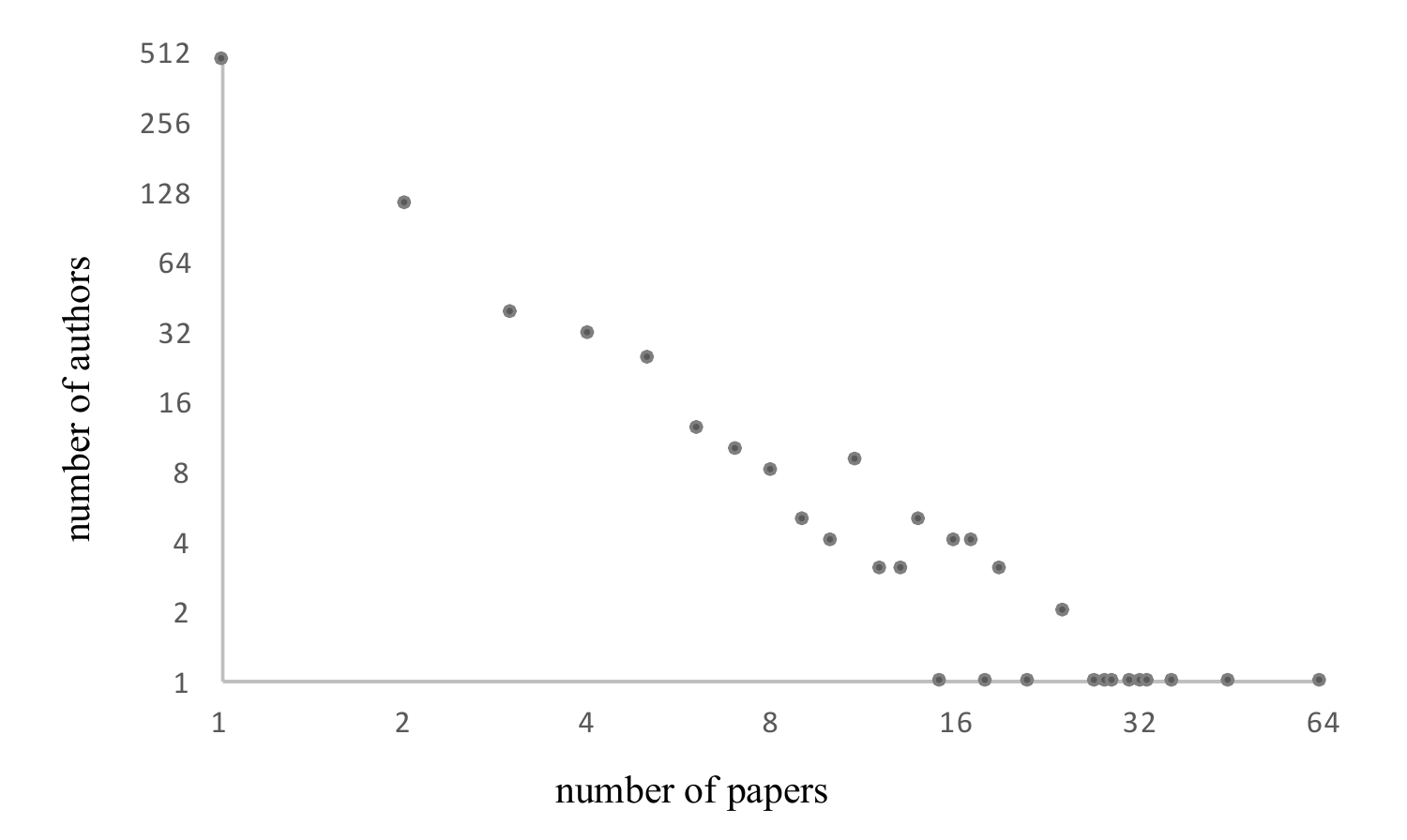}

\caption{\label{Distribution of publications}Distribution of paper publication
on a log-log scale: 1981-2010. A small number of authors produce a
disproportionate fraction of papers written in the field. }
\end{figure}

Over this time period, there are a total of 760 unique authors.  Two authors have been active in our data set over the entire 30 year
period. Authors publish an average one paper per year. However, not
all authors produce the same number of papers. Figure \ref{Distribution of publications}
shows a log-log plot of the distribution of papers across authors.
This distribution is quite skewed, meaning that a small number of
authors publish a disproportionate number of papers---the most prolific
20\% of authors in the field have written over 65\% of the papers.
A similar pattern can be seen in many other academic fields\cite{Newman2001a,grossman2002evolution,Moody2004}.

\begin{figure}
\includegraphics[width=\columnwidth]{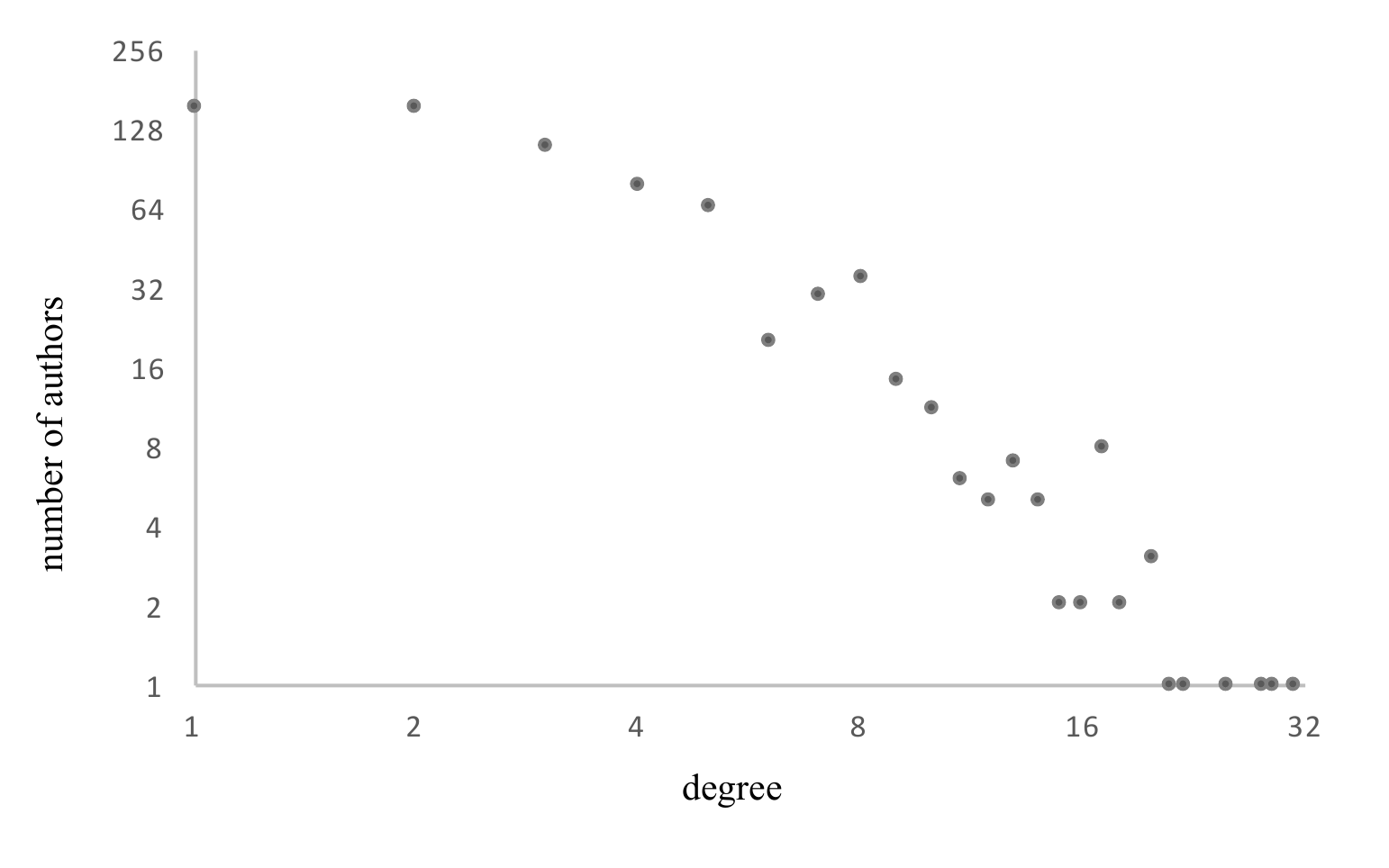}

\caption{\label{Degree Distribution}Degree distribution for the PER network
on a log-log plot: 1981-2010. Degree on the coauthorship network is
the number of coauthors an author has. A small number of authors publish
with a large fraction of the field, while most authors publish with
very few.}
\end{figure}

\subsection{Coauthorship behavior}

The coauthorship network for PER is much as one would expect for a field of academic research. In PER, as in many academic areas, there is wide variation in coauthorship behavior among authors. A handful of researchers have a large number of coauthors, while most researchers have very few. The average researcher has 4.1 coauthors, but three authors (0.3\%) have more than 30 coauthors, 12 authors (1.2\%) have more than 20 coauthors, and 60 authors (6.3\%) have more than 10 coauthors. (see Figure \ref{Degree Distribution}). A similar pattern can be seen in many other academic fields, including physics\cite{Newman2001a,Newman2001PNAS}, biology\cite{Newman2001a,Newman2001PNAS,powell2005network}, math\cite{grossman2002evolution}, neuroscience\cite{barabasi2002evolution} economics\cite{Goyal2006}, sociology\cite{Moody2004}, and business\cite{Acedo2006a}.

\begin{figure}
\includegraphics[width=\columnwidth]{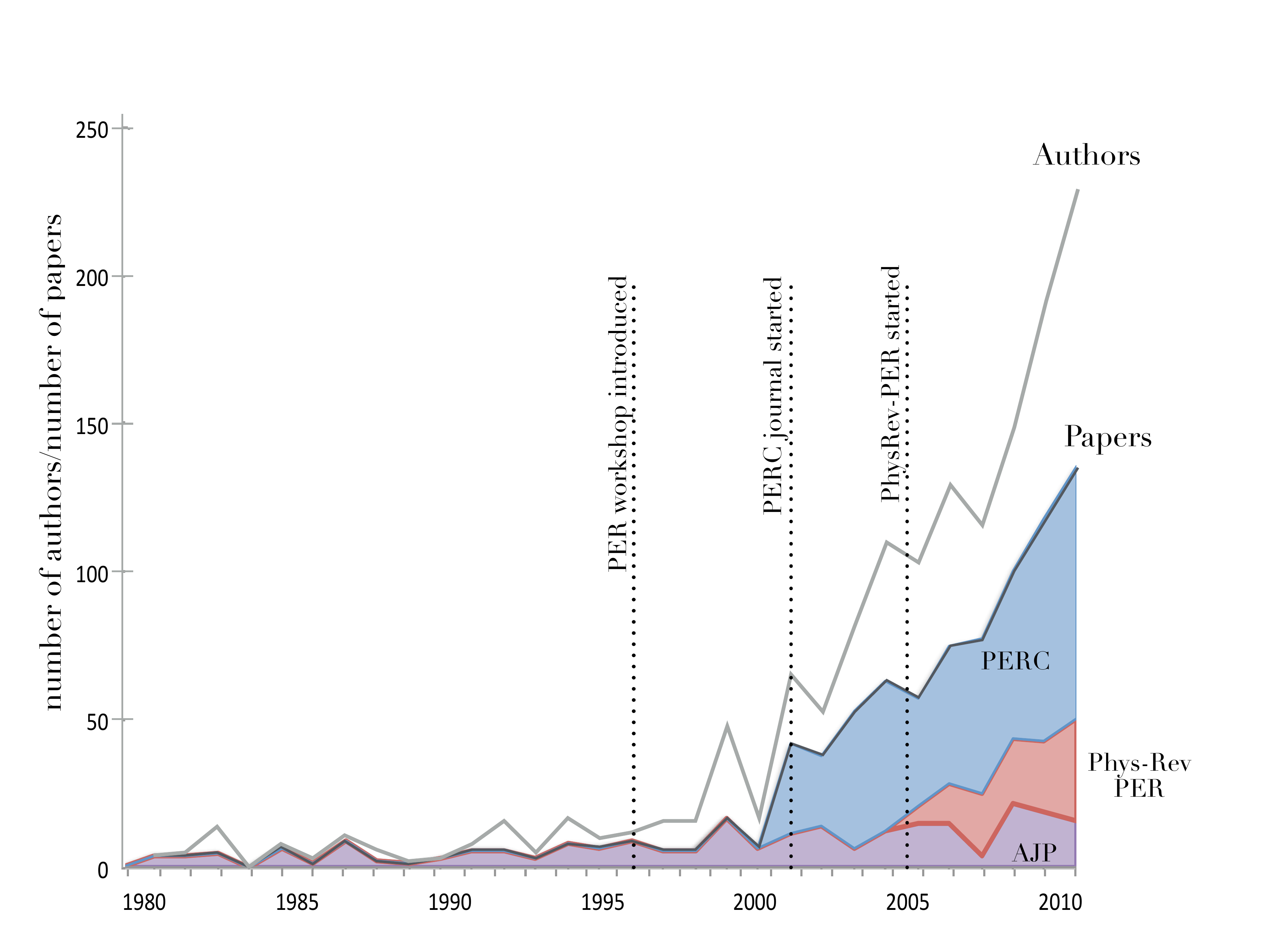}

\caption{\label{Number of papers Number of authors}The number of PER authors,
and the number of PER papers. Papers are broken down by journal. Important
events in the field are highlighted.}
\end{figure}

\subsection{Central members}
In addition to the number of papers for each person, both degree centrality and eigenvector centrality have been proposed as measures of prominence in a community. Table \ref{Top Authors} lists the top 5 authors in the community according to number of papers, number of coauthors, and eigenvector centrality.

Recent work has suggested that eigenvector centrality may be a better measure of prominence in a community than degree. In particular, there is some indication that individuals with high eigenvector centrality are more influential when it comes to disseminating information \cite{banerjee2013diffusion}. In the context of academic production, one might also argue that degree does not capture the relationship between advisors and their graduate students: authors who primarily coauthor with graduate students who then leave the field are less prominent than authors who mentor successful graduate students, and those who work with other giants in the field. In Figure \ref{Coauthorship network 1981-2010}, the nodes are sized by the number of coauthors and colored by eigenvector centrality. The nodes with the highest eigenvector centrality are not, generally speaking, the nodes with the highest degree. 

Interestingly, the top members of the eigenvector centrality group are all senior or former members of the University of Colorado at Boulder (PER@C) group, while the top members of the number of coauthors and number of publications groups come from a much wider distribution of research groups. The high eigenvector centrality of faculty at the University of Colorado researchers is partially attributable to their graduate students and postdocs: alumni from this program are unusually successful as they proceed in their careers. This lends credence to the idea that the difference between eigenvector centrality and degree centrality reflects differences in the success of advisors in producing quality graduate students and postdocs.

\begin{table*}
\begin{tabular}{|c|c|c|c|c|c|}
\hline 
number of publications &  & number of coauthors &  & eigenvector centrality & \tabularnewline
\hline 
\hline 
Noah D. Finkelstein & 69 & Sanjay Rebello & 39 & K.K. Perkins & .47\tabularnewline
\hline 
Chandralekha Singh & 65 & Noah D. Finkelstein & 32 & Noah D. Finkelstein & .44\tabularnewline
\hline 
Sanjay Rebello & 54 & Lei Bao & 31 & S.J. Pollock & .43\tabularnewline
\hline 
Charles Henderson & 41 & K.K. Perkins & 28 & Wendy K. Adams & .26\tabularnewline
\hline 
S.J. Pollock & 40 & Robert Beichner & 22 & Carl E. Wieman & .25\tabularnewline
\hline 
\end{tabular}

\caption{\label{Top Authors}Top authors by 1) number of papers published 2)
number of coauthors (degree in the coauthorship network) 3) coauthor
prominance (eigenvector centrality in the coauthorship network) }

\end{table*}

\section{Community Growth}

As can be seen in Figure \ref{Number of papers Number of authors},
the field of PER has grown dramatically in the past 30 years, with
most of that growth in our data set occurring in the last ten years. Initially, the
number of papers grows slowly but in the early 2000s, it explodes.
The growth in the number of authors is very similar, indicating that
there is a growth in the overall size of the field, rather than simply
an increase in the average number of papers written per person. 

\begin{figure*}
\includegraphics[width=\textwidth]{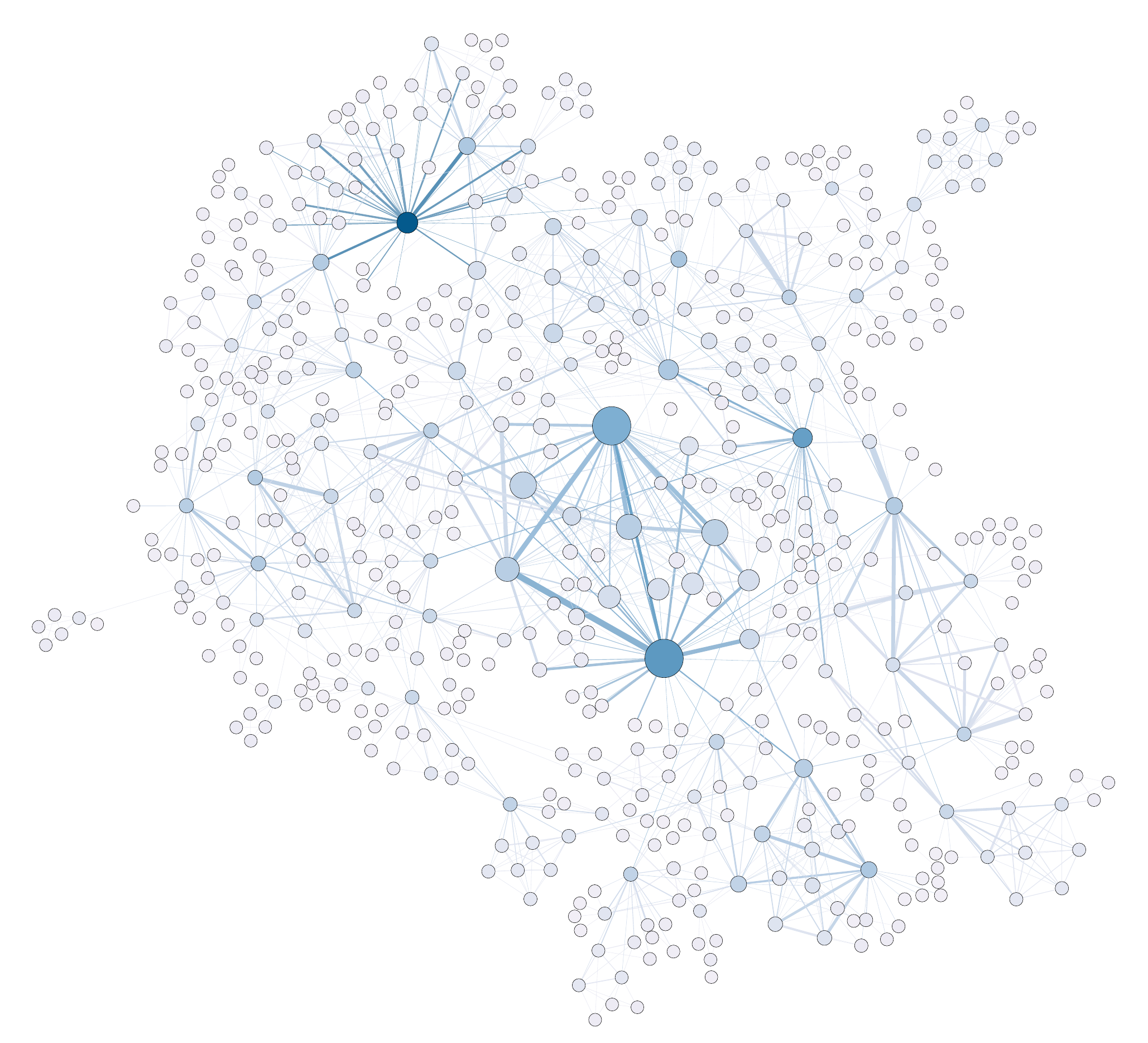}

\caption{\label{Coauthorship network 1981-2010}The largest connected component of the
PER coauthorship network (1981-2010). The color represents degree
and the size represents eigenvector centrality.}
\end{figure*}

In addition to the overall growth, this period also sees a dramatic
change in the pattern of collaborative interactions within the field,
as tracked through the coauthorship network. Figure \ref{Coauthorship network 1981-2010}
shows a representation of the LCC of the PER coauthorship
network, aggregated over the entire time period. 

\subsection{The emergence of a community}

This aggregate picture of the community is largely in line with what we see in other scientific fields. However, one of the more interesting aspects of these data is the opportunity to look at the development of the community over time, particularly as it moves from the very earliest stages into a more mature community. In several ways, the evolution of this network is similar to that seen in longitudinal studies of mature fields. But we also observe several features that appear to be unique to early-stage collaborative communities.

\begin{figure}
\includegraphics[height=0.8\paperheight]{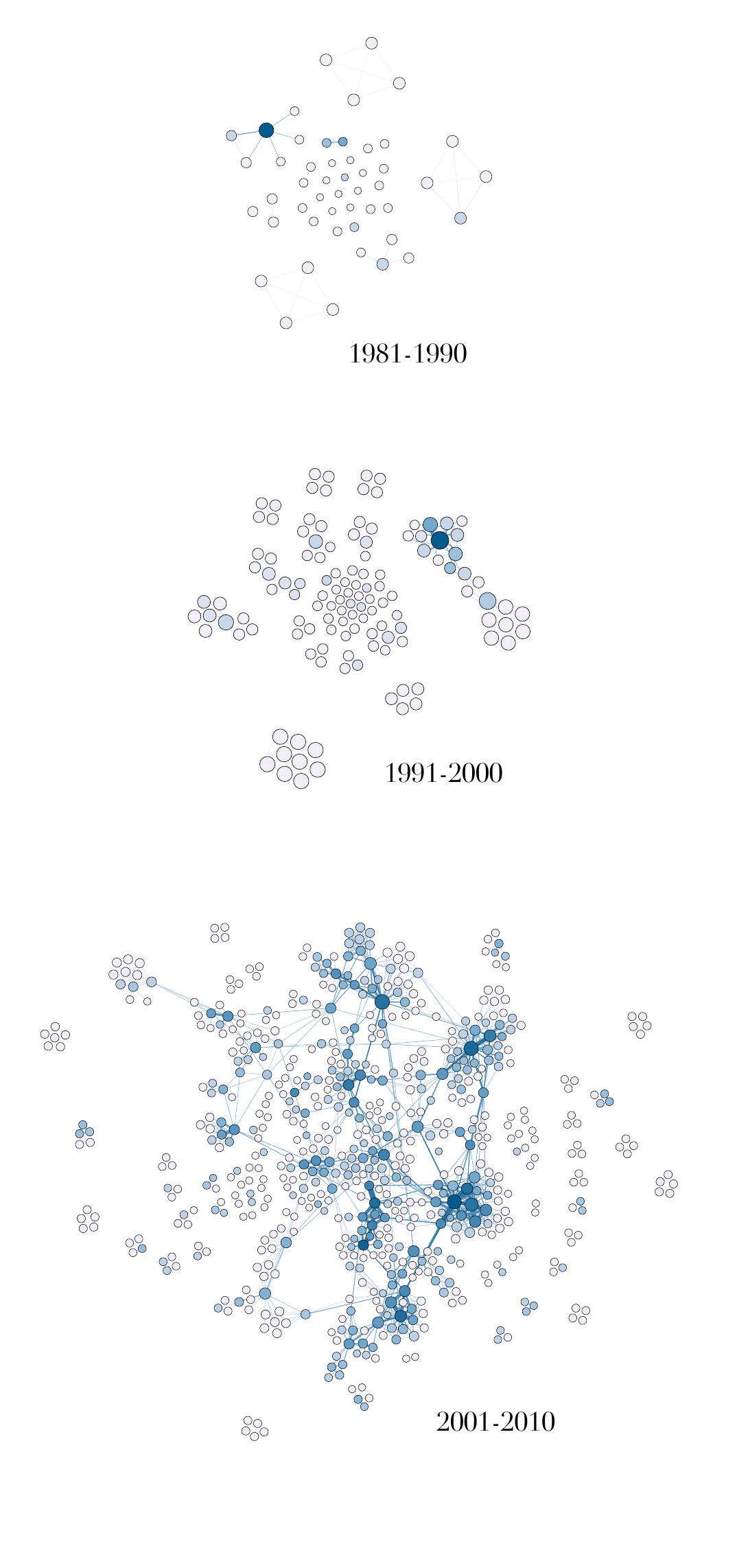}

\caption{\label{Evolution of network}The shape of the collaboration network
in the 1980s, 1990s, and 2000s. }
\end{figure}

Figure \ref{Evolution of network} shows a network visualization of the PER community over three different time periods: the 1980s, 1990s, and 2000s. These networks illustrate how the shape of the PER community has changed over the last 30 years. During the 1980s, the authors in the field labored more or less independently. This was also true through the 1990s. But in the early-to-late 2000s, the network consolidates into a single community with a cohesive core.

We can see this consolidation of the community by looking at the number of people in the LCC. In most of the mature research communities that have been studied, the largest connected component contains well over half of the researchers, ranging from 57\% in computer science to 93\% in the biomedical fields (see Table \ref{Percent in Largest Connected Component} for a list of established results \footnote{Biomedical, astrophysics, condensed matter physics, and high-energy theory: \cite{Newman2001PNAS} Sociology: \cite{Moody2004} Management: \cite{Acedo2006a}}.) We can think of the nodes in the LCC as the core of the community, and when that core community contains a large fraction of the researchers in the field, the community hangs together as a single, cohesive unit.

\begin{table}
\begin{tabular}{|c|c|}
\hline 
field & frac in LCC\tabularnewline
\hline 
\hline 
biomedical (1995-1999) & 93\%\tabularnewline
\hline 
astrophysics (1995-1999) & 89\%\tabularnewline
\hline 
condensed matter physics (1995-1999) & 85\%\tabularnewline
\hline 
high-energy theory (1995-1999) & 71\%\tabularnewline
\hline 
management (1980-2003) & 45\%\tabularnewline
\hline 
sociology (1963-1999) & 53\%\tabularnewline
\hline 
PER (80s and 90s) & 12\%\tabularnewline
\hline 
PER (2000s) & 68\%\tabularnewline
\hline 
\end{tabular}

\caption{\label{Percent in Largest Connected Component}The fraction of researchers in
the largest connected component in various established research communities,
compared with PER. }

\end{table}

In contrast, the LCC in the early-stage PER network---formed using papers from the 80s and 90s---contains only 12\% of the researchers writing PER papers. This suggests that in those early years, PER was not a cohesive collaborative community, such as those seen in more established fields. However, the later-stage network does exhibit a cohesive core: the LCC in the network constructed using papers from the 2000s contains 68\% of the researchers in the community.

\begin{figure}
\includegraphics[width=\columnwidth]{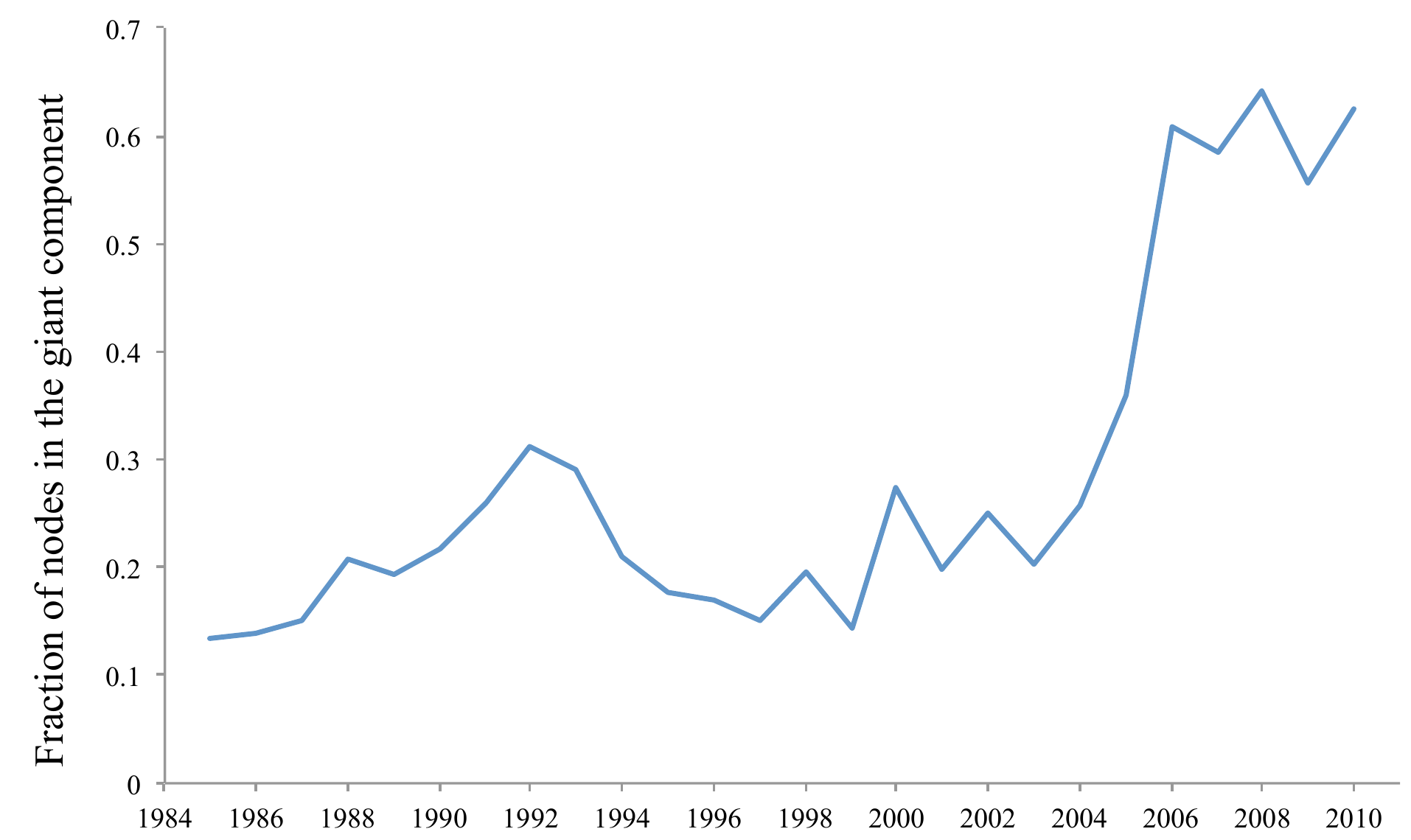}

\caption{\label{Fraction in LCC}The fraction of nodes in the largest connected
component of the network. The emergence of a cohesive core occurs
in the period from 2004-2005.}
\end{figure}

Interestingly, the transition to a cohesive community does not occur gradually. There is a clear point at which the largest connected component starts to dominate the network. Figure \ref{Fraction in LCC} shows the fraction of the PER community that is in the largest connected component, using networks generated from papers within a five-year moving window. For any date, a five-year moving window starts two years prior, and ends two years after. Thus, the size of the LCC in 1995 is calculated by averaging the sizes in 1993-1997. We use five-year windows because the earliest networks are generated by a very small set of papers. The results are similar for different window sizes.

In 2004-2005, we see a discrete jump in the fraction of nodes that are connected, marking the transition from isolated islands of researchers to a cohesive core. We will examine this jump further below.

\subsection{More collaborative effort}

\begin{figure}
\includegraphics[width=\columnwidth]{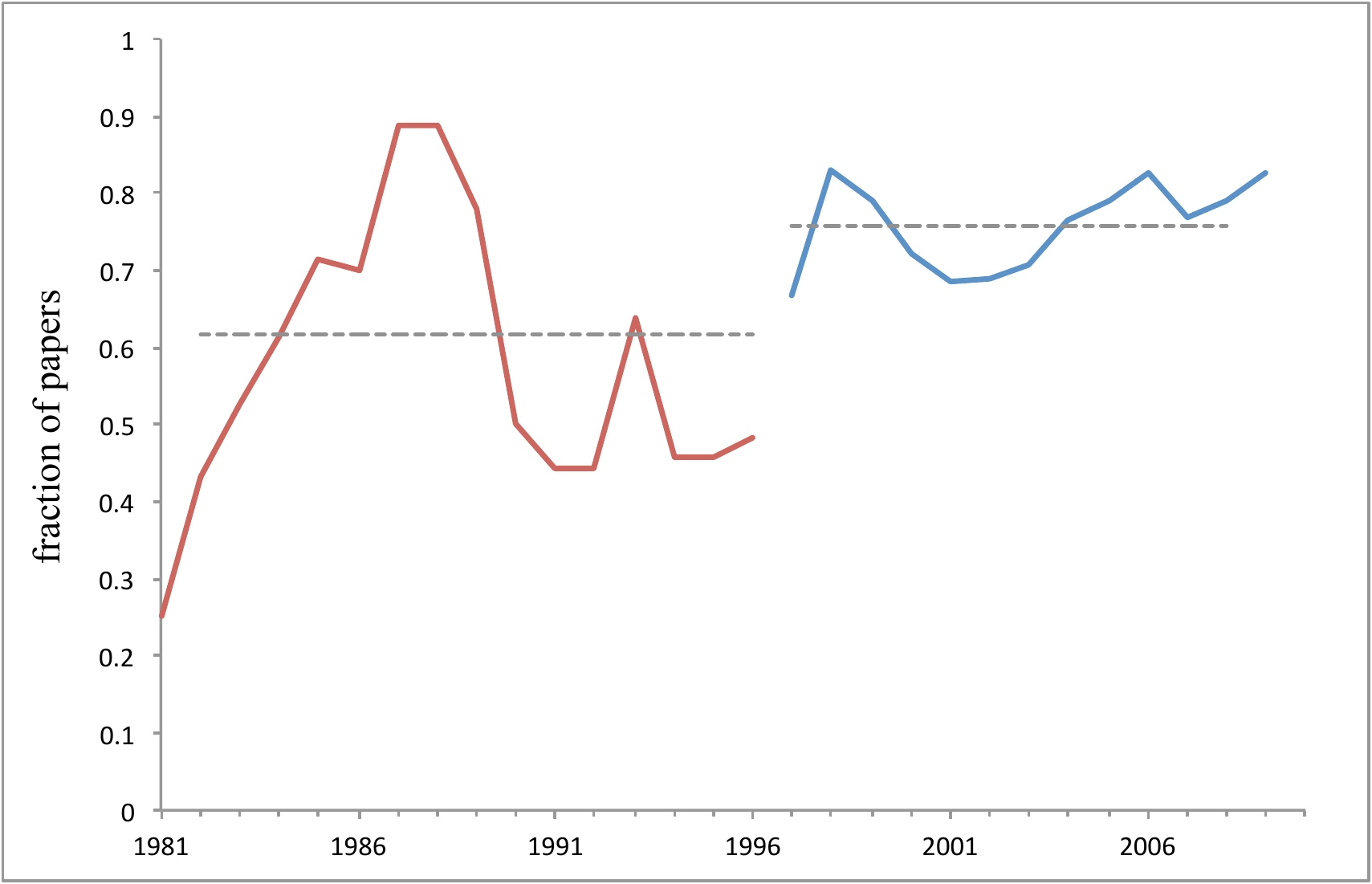}

\caption{\label{Fraction sole-authored}The fraction of papers with multiple
authors. The fraction of papers coauthored exhibits a discrete jump
in 1997. The dotted lines indicate the average rate of coauthorship
before and after this break.}
\end{figure}

Over time, PER has become an increasingly collaborative field. In
the early 1980s, around half of the papers in the field had a single
author. By 2010, only a quarter were solo-authored. Many fields have
seen a similar increase in collaboration over time. In the field of
sociology, for example, the average fraction of coauthored papers
rose from \textasciitilde{}20\% in 1963 to \textasciitilde{}40\% in
1999\cite{Moody2004}. Similar long-term trends have been observed
in other established fields, such as information science\cite{cunningham1997authorship},
mathematics\cite{grossman2002evolution}, and sociology\cite{Moody2004}.
However, whereas other fields have experienced a slow, steady increase
in coauthorship rates, the change in PER did not come about gradually.
Rather, there was a discrete change in collaborative behavior in 1997-1998.
Figure \ref{Fraction sole-authored} shows the time series of the
fraction of papers coauthored, with mean values before and after 1997.
Before 1997, 53\% of papers were coauthored. After 1997, the propensity
to collaborate jumped to nearly 80\%.

\begin{figure}
\includegraphics[width=\columnwidth]{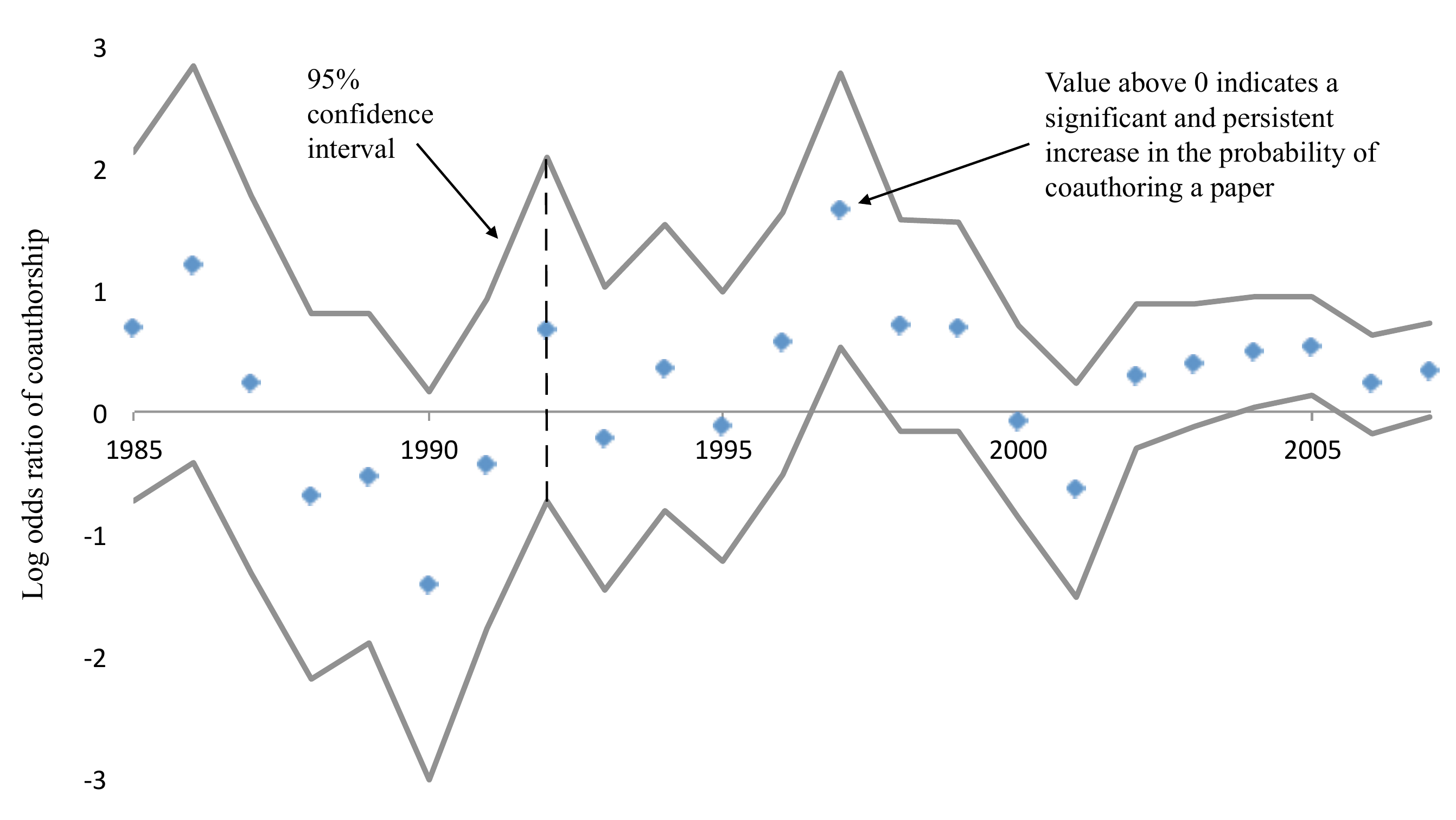}

\caption{\label{1997 is unique}The change in the log odds of a paper in AJP
being coauthored in the four years before and after a given point.
Grey lines indicate the 95\% confidence interval on the point estimate
of the log odds ratio. A log odds ratio above 0 indicates a persistent
increase in the odds of a paper being coauthored. This occurs in only
one time period for this journal: 1997. }
\end{figure}

Using a log odds ratio, we can show that this discrete change in the propensity to collaborate is significant, persistent, and unique. Here, we look at the data from AJP alone, so as to keep the venue for publication consistent throughout. The change at this breakpoint is still significant when all three journals are included. We compare the odds of coauthoring a paper in the 4 years before and after each year. Suppose $a_{1}$ and $b_{1}$ are the number of coauthored and single-authored papers in the four years prior to a given year, and $a_{2}$ and $b_{2}$ are the number of coauthored and single-authored papers in the four years after that given year. The log odds ratio is then $y=\left(log(a_{2})-log(b_{2})\right)-\left(log(a_{1})-log(b_{1})\right)$ Figure \ref{1997 is unique} shows the change in the log odds ratio over time, with a 95\% confidence interval. The 95\% confidence interval for the point estimate of the log odds ratio is $ln(y)\pm1.96*ste(ln(y))$, where $ste(ln(y))=\sqrt{\frac{1}{a_{1}}+\frac{1}{b_{1}}+\frac{1}{a_{2}}+\frac{1}{b_{2}}}$.

A log odds ratio above 0 means that papers are more likely to be coauthored in the four years after the break than they are in the four years before the break; in 1997 the 95\% confidence interval excludes 0 and is statistically significant. Note that while the choice of a 4 year window is arbitrary, the break in 1997 remains significant for other window-sizes.  Because the change has to continue through the post-break window, the positive log odds ratio necessarily indicates a persistent change in the probability of coauthoring a paper. Moreover, this period of dramatically increased collaboration is relatively unique in the post-1981 history of PER.  

There is only one other significantly positive log odds ratio, in 2005. This break point occurs a few years after the introduction of the PERC Proceedings and it is coincident with the introduction of the PhysRev-PER journal. The first few years of both PERC Proceedings and PhysRev-PER were unusually collaborative, which undoubtedly generates the significant break point in 2005. However, in contrast with the break point in 1997, this appears to be a temporary effect.

The intensive margin---the number of coauthors per paper---provides some additional insight into the mechanisms behind the observed increase in collaboration. This measure of collaborative effort has also risen in a wide range of other academic fields, including physics\cite{BARABASI2002}, mathematics\cite{Grossman1995}, sociology\cite{Moody2004}, management science\cite{Acedo2006a}, and economics\cite{Goyal2006}. Looking at the overall number of coauthors per paper, there initially appears to be a similar, though less dramatic trend in PER: the number of authors per paper rises from 2.2 in the 80s and 90s to 2.5 in the 2000s. However, when we condition on a paper being coauthored at all, the trend disappears: the average coauthored paper had 3.0 authors in the 80s and 90s and 2.9 authors in the 00s. This suggests that in the case of PER, the change in behavior was regarding whether to coauthor or not, rather than in the number of authors to bring onto a project.

This raises the question of where this sudden move towards collaboration came from. It is worth noting that 1997 is coincident with the introduction of a field-specific conference---PERC---which developed from grass-roots efforts at Kansas State University, the University of Nebraska-Lincoln, and the University of Maryland, and was later fully recognized as an extension to the summer meeting of the American Association of Physics Teachers. PERC attendance grew to approximately 300 annual registrants by the end of our data collection period. At this conference, researchers---including graduate students and incumbents---could meet and foster new collaborative relationships. Without information on conference attendance, it is impossible to tell whether the introduction of the conference facilitated this increase in collaborative effort. However, the fact that collaboration has been higher in the PERC-era suggests that the conference may have been a factor.

Another possible contributing factor is the increased use of email during this period, which lowered the costs of remote collaboration, perhaps prompting increased probability of coauthorship. The field used email and other internet communications to foster specific collaborations as well as develop online communities such as the graduate students' mailing list (then ``Graduate Students in Physics Education'' (GSPER), now ``PER Consortium of Graduate Students'' (PERCoGS)). Technological advances in the collection and distribution of raw data (survey responses, video files) also made it easier to collect, share, and analyze data among geographically-distant collaborators. However, if these technological factors were truly dominant, we would expect to see a similar discrete jump in coauthorship in other fields over roughly the same time period---especially in other early-adopter fields such as high-energy physics. The fact that we do not suggests that increased internet use is unlikely to be the dominant source of the observed change.

\subsection{Increasing prevalence of within-group links}

\begin{figure}
\includegraphics[width=\columnwidth]{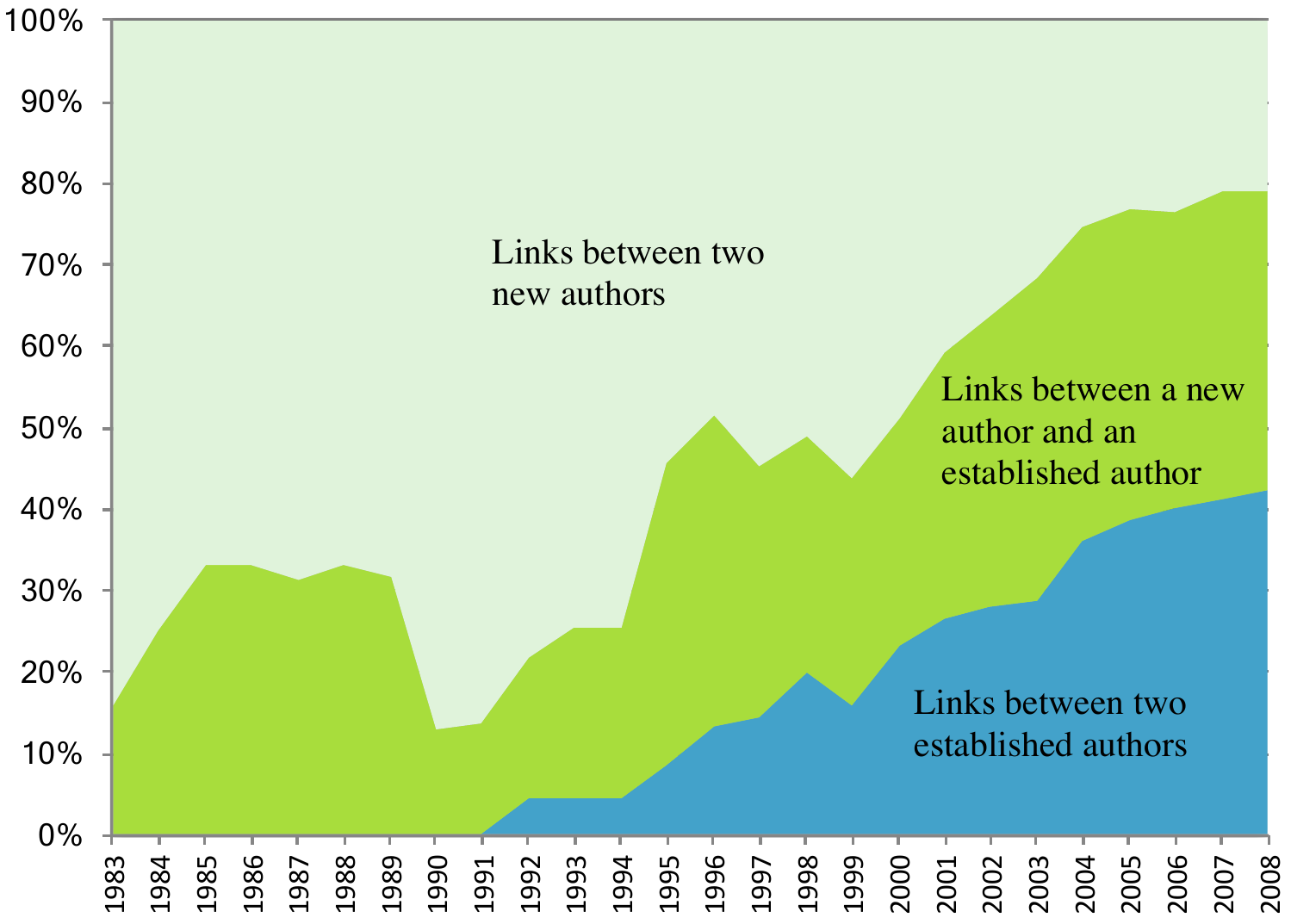}

\caption{\label{New authors}Fraction of PER coauthor relationships that are
between existing members of the field (bottom, blue), between new members of the
field (middle, dark green), and between existing and new members (top, pale green). }
\end{figure}

In addition to an increase in the propensity to collaborate, we also see a change in the pattern of collaboration. Over the 30 years covered by the data, PER researchers shift from working with researchers from outside the community to researchers who are already active in the field. Figure \ref{New authors} shows the fraction of coauthorship ties that are between authors who have already published in a PER journal, the fraction that are between new authors, and the fraction that include both a new author and an incumbent author. Early on, existing members of the PER community tend to work with researchers new to the field. This trend persists for a surprisingly long time: 15 years into the observed data, fewer than 10\% of the coauthorship links are between researchers who are already established in the field. But as the community ages, established community members begin to work with other established community members. By the late 2000s, about half of the links are between researchers who are already active in the field. 

\begin{figure}
\includegraphics[width=\columnwidth]{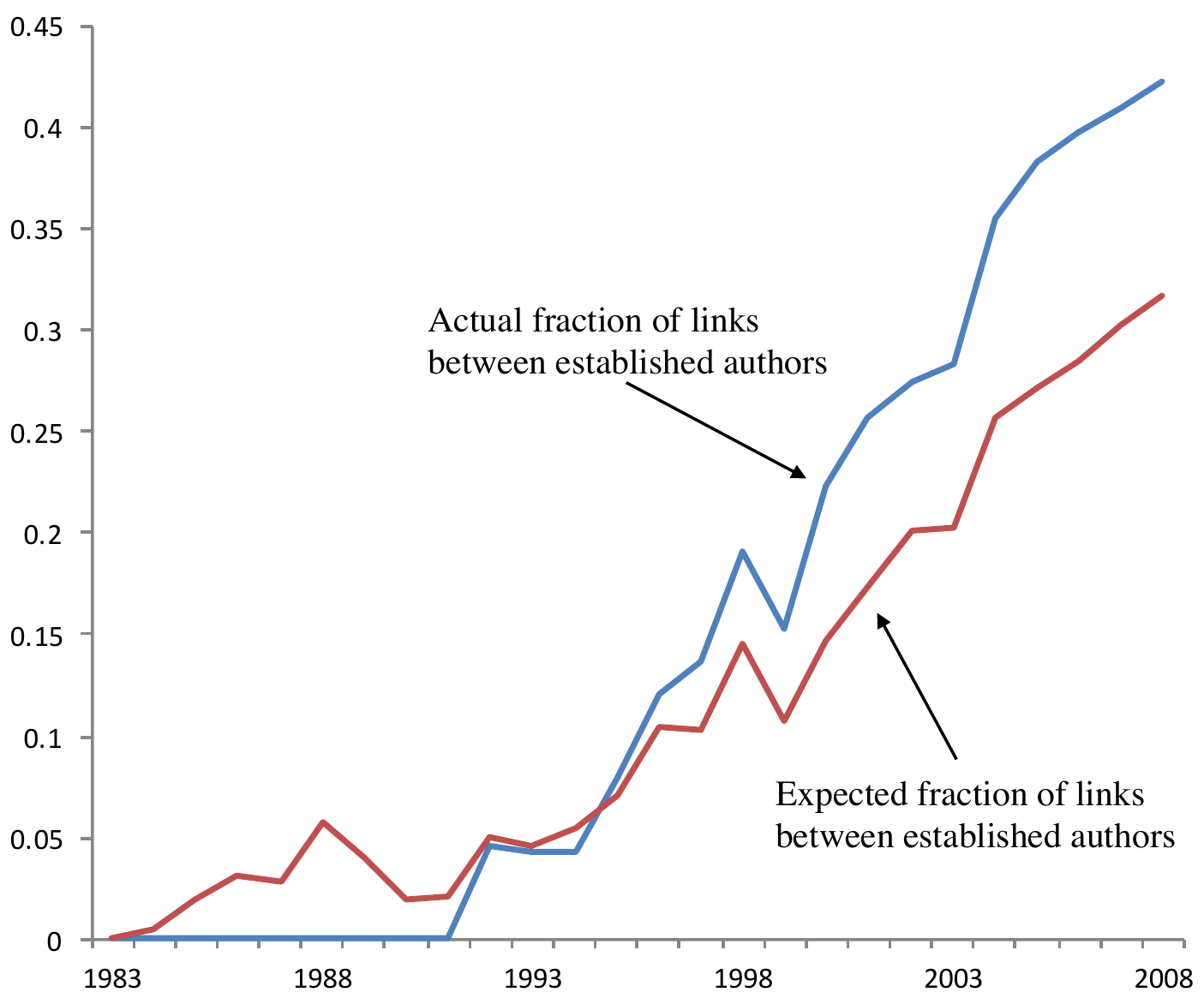}

\caption{\label{Excepted ii links}The fraction of coauthorships that occur
between members of the community (blue), compared with the number that would
be expected if linking behavior were random (red). }
\end{figure}

Of course, this trend need not be due to a true shift in linking behavior, because as the field matures, there are an increasing number of incumbents, which provides more ample opportunities for interaction between them. We can account for the aging of the community by comparing the fraction of links that are between incumbent authors to the fraction that would be expected if links between those authors were made at random. This is illustrated in Figure \ref{Excepted ii links}. In the early years of the field, researchers were actually less likely to coauthor with established PER researchers than would be expected. But in the mid-to-late 90s, we see a reversal in that trend, as established researchers become increasingly likely to coauthor. By mid-to-late 2000s, established researchers have shifted towards disproportionately collaborating with researchers who are established in the field.

While the increase in collaborative propensity is a trend found in many different fields, this shift towards within-group collaboration has not, to our knowledge, been previously observed in other academic contexts. There are many overlapping factors which could contribute to the observed changes in coauthorship behavior. It seems likely that it is a trend unique to early-stage academic communities. Before the field becomes established as a cohesive community, established researchers coauthor with a rotating cast of colleagues new to the field, most of whom never write another paper in that area. As the field develops and grows, more researchers are able to specialize in that area, allowing for an increase in coauthorship between established members of the field.

The appearance of field-specific graduate programs is also a likely factor. Graduate students provide an important link between established researchers, because they maintain connections with their old institutions while establishing connections at new institutions. This geographic movement provides an opportunity for established authors to work with each other over time, even when they are no longer at the same institutions. In the 1990s, graduate programs in PER expanded even as graduate programs in physics as a whole saw declines in admissions\cite{AIP2015FirstYear}. In the network, these authors appear as bridges between multiple research groups, with many links to multiple researchers at two different institutions which otherwise do not have strong direct links among them. Notable cases of inter-institutional links include David Brookes and Elizabeth Gire.  Other established researchers are only strongly linked to one group.  As the field grows there are also likely to be more opportunities for researchers to take sabbaticals in distant locations, increasing extra-institutional and international links within the community. Researchers may also form remote collaborations based on mutual interests, as in the cases of Melissa Dancy and Charles Henderson or Eric Brewe and Rachel Scherr, even if their home institutions always geographically distant.  

\section{Discussion}

The changes we see in Figure \ref{Evolution of network} and Figure \ref{Fraction in LCC} were sudden and dramatic. The transition from individual researchers to an interconnected whole is a clear indication of the emergence of a collaborative community, and thus it is natural to ask which factors may have contributed to that change. In this section, we consider the roles of growth, increased collaboration, and increased within-group collaboration in the development of the community.

The first, and simplest explanation for the consolidation of the network is growth:  as more papers are written, there are more opportunities for coauthorship. However, this explanation does not fit the data.  We modeled expected network density to test this explanation.

Assuming that linking decisions are random (thus eliminating factors related to linking behavior), the expected density of the network at time t is approximately \begin{align}
 \frac{expected\ number\ connections}{total\ connections\ possible}=\frac{N_{p}\left(p_{c}\right)}{N_{p}\left(N_{p}-1\right)} 
 \end{align} 
 where $N_{p}$ is the number of authors and $p_{c}$ is the probability that a paper is co-authored.   This approximation assumes that all papers that are co-authored are written by two authors or solo-authored. This is a reasonable approximation for our data, because the median paper is written by between one and two individuals, depending on the year.

If growth in the number of papers were the sole factor in the emergence of the cohesive core, then $p_{c}$ would be constant, and density would decrease over time, which would actually inhibit the formation of a large connected component that dominates the graph. Thus, growth alone cannot explain the emergence of a cohesive core.

A second possibility is that this increased cohesion is a result of one of the observed changes in linking behavior: either the increase in the number of authors per paper, or the shift toward within-group coauthorship. The increase in coauthorship rates seen in Figures \ref{Fraction sole-authored} and \ref{1997 is unique} means more links, which would provide more opportunities for isolated parts of the network to come together.

Another possibility is that the consolidation of the community is a result of not the overall amount of collaboration, but rather the choice of who to collaborate with: i.e. the trend towards existing members of the community authoring papers together. Isolated groups of existing authors could be bound together by ``long distance'' links between existing community members, which would lead to the emergence of the cohesive core.

It is also worth noting that many of the behavior and structural changes we observe are coincident with changes in the institutional structures in the community. The uptick in collaboration occurs around the same time as the ``interval day'' meeting in 1996, a precursor to the PERC. Moreover, this is coincident with the introduction of the graduate student mailing list (originally GSPER, now PERCoGS), which undoubtedly increased inter-departmental communication, and thus community cohesion.

This period also saw increases to PER-specific funding at the National Science Foundation, increasing both the opportunities for newcomers and the possible projects for them to work on.  Concurrently, an explosion of research-based teaching methods\cite{Meltzer2012} and research-based assessments\cite{Madsen2017RBAI} allowed many faculty access to products of PER, opening possibilities for future research and collaboration.

\section{Conclusion}

The case of PER provides a useful look at the
early stages in the development of an academic community. The community
grows, and with that growth comes an increased propensity to collaborate
and an increased reliance on collaboration within the community. The
introduction of field-specific conference also provided a valuable
forum for researchers to develop a sense of community and shared purpose.
The result is the emergence of a new, cohesive core to the coauthorship
network---a true academic community.

The interactions between field growth, behavioral change, and structural change are complex, and it is impossible to completely tease out the effects of each factor. It is likely their effects interact strongly with each other, forming feedback loops to increase the strength and interconnections within the community.

\begin{acknowledgments}

The authors thank Lyle Barbato, who provided the PERC Proceedings data, and Steve Kanim, who provided the AJP data.  Several community-minded individuals contributed background information and potential real-world correlates to this paper, including Michael Wittmann, Dean Zollman, Michael Loverude, and two anonymous reviewers.  
Portions of this research were supported by the KSU Physics Department and by the Tepper School of Business, Carnegie Mellon University.

\end{acknowledgments}


\begin{thebibliography}{25}%
\makeatletter
\providecommand \@ifxundefined [1]{%
 \@ifx{#1\undefined}
}%
\providecommand \@ifnum [1]{%
 \ifnum #1\expandafter \@firstoftwo
 \else \expandafter \@secondoftwo
 \fi
}%
\providecommand \@ifx [1]{%
 \ifx #1\expandafter \@firstoftwo
 \else \expandafter \@secondoftwo
 \fi
}%
\providecommand \natexlab [1]{#1}%
\providecommand \enquote  [1]{``#1''}%
\providecommand \bibnamefont  [1]{#1}%
\providecommand \bibfnamefont [1]{#1}%
\providecommand \citenamefont [1]{#1}%
\providecommand \href@noop [0]{\@secondoftwo}%
\providecommand \href [0]{\begingroup \@sanitize@url \@href}%
\providecommand \@href[1]{\@@startlink{#1}\@@href}%
\providecommand \@@href[1]{\endgroup#1\@@endlink}%
\providecommand \@sanitize@url [0]{\catcode `\\12\catcode `\$12\catcode
  `\&12\catcode `\#12\catcode `\^12\catcode `\_12\catcode `\%12\relax}%
\providecommand \@@startlink[1]{}%
\providecommand \@@endlink[0]{}%
\providecommand \url  [0]{\begingroup\@sanitize@url \@url }%
\providecommand \@url [1]{\endgroup\@href {#1}{\urlprefix }}%
\providecommand \urlprefix  [0]{URL }%
\providecommand \Eprint [0]{\href }%
\providecommand \doibase [0]{http://dx.doi.org/}%
\providecommand \selectlanguage [0]{\@gobble}%
\providecommand \bibinfo  [0]{\@secondoftwo}%
\providecommand \bibfield  [0]{\@secondoftwo}%
\providecommand \translation [1]{[#1]}%
\providecommand \BibitemOpen [0]{}%
\providecommand \bibitemStop [0]{}%
\providecommand \bibitemNoStop [0]{.\EOS\space}%
\providecommand \EOS [0]{\spacefactor3000\relax}%
\providecommand \BibitemShut  [1]{\csname bibitem#1\endcsname}%
\let\auto@bib@innerbib\@empty
\bibitem [{\citenamefont {Traweek}(1988)}]{Traweek1988}%
  \BibitemOpen
  \bibfield  {author} {\bibinfo {author} {\bibfnamefont {S}~\bibnamefont
  {Traweek}},\ }\href@noop {} {\emph {\bibinfo {title} {{Beamtimes and
  lifetimes}}}}\ (\bibinfo  {publisher} {Harvard University Press},\ \bibinfo
  {address} {Cambridge, MA},\ \bibinfo {year} {1988})\BibitemShut {NoStop}%
\bibitem [{\citenamefont {Grossman}\ and\ \citenamefont
  {Ion}(1995)}]{Grossman1995}%
  \BibitemOpen
  \bibfield  {author} {\bibinfo {author} {\bibfnamefont {Jerrold~W}\
  \bibnamefont {Grossman}}\ and\ \bibinfo {author} {\bibfnamefont {Patrick
  D~F}\ \bibnamefont {Ion}},\ }\bibfield  {title} {\enquote {\bibinfo {title}
  {{On a Portion of the Well-Known Collaboration Graph}},}\ }\href
  {http://citeseerx.ist.psu.edu/viewdoc/download?doi=10.1.1.35.4242\&amp;rep=rep1\&amp;type=ps}
  {\bibfield  {journal} {\bibinfo  {journal} {Congressus Numerantium}\ }\textbf
  {\bibinfo {volume} {108}},\ \bibinfo {pages} {129--131} (\bibinfo {year}
  {1995})}\BibitemShut {NoStop}%
\bibitem [{\citenamefont {Newman}(2001{\natexlab{a}})}]{Newman2001a}%
  \BibitemOpen
  \bibfield  {author} {\bibinfo {author} {\bibfnamefont {M.E.J.}\ \bibnamefont
  {Newman}},\ }\bibfield  {title} {\enquote {\bibinfo {title} {{Scientific
  collaboration networks. I. Network construction and fundamental results}},}\
  }\href {\doibase 10.1103/PhysRevE.64.016131} {\bibfield  {journal} {\bibinfo
  {journal} {Physical Review E}\ }\textbf {\bibinfo {volume} {64}},\ \bibinfo
  {pages} {1--8} (\bibinfo {year} {2001}{\natexlab{a}})}\BibitemShut {NoStop}%
\bibitem [{\citenamefont {Newman}(2001{\natexlab{b}})}]{newman2001b}%
  \BibitemOpen
  \bibfield  {author} {\bibinfo {author} {\bibfnamefont {Mark~EJ}\ \bibnamefont
  {Newman}},\ }\bibfield  {title} {\enquote {\bibinfo {title} {Scientific
  collaboration networks. ii. shortest paths, weighted networks, and
  centrality},}\ }\href@noop {} {\bibfield  {journal} {\bibinfo  {journal}
  {Physical review E}\ }\textbf {\bibinfo {volume} {64}},\ \bibinfo {pages}
  {016132} (\bibinfo {year} {2001}{\natexlab{b}})}\BibitemShut {NoStop}%
\bibitem [{\citenamefont {Newman}(2001{\natexlab{c}})}]{Newman2001PNAS}%
  \BibitemOpen
  \bibfield  {author} {\bibinfo {author} {\bibfnamefont {M~E~J}\ \bibnamefont
  {Newman}},\ }\bibfield  {title} {\enquote {\bibinfo {title} {{The structure
  of scientific collaboration networks}},}\ }\href
  {http://arxiv.org/abs/cond-mat/0007214} {\bibfield  {journal} {\bibinfo
  {journal} {Proceedings of the National Academy of Sciences of the United
  States of America}\ }\textbf {\bibinfo {volume} {98}},\ \bibinfo {pages} {7}
  (\bibinfo {year} {2001}{\natexlab{c}})}\BibitemShut {NoStop}%
\bibitem [{\citenamefont {Newman}(2004)}]{newman2004coauthorship}%
  \BibitemOpen
  \bibfield  {author} {\bibinfo {author} {\bibfnamefont {Mark~EJ}\ \bibnamefont
  {Newman}},\ }\bibfield  {title} {\enquote {\bibinfo {title} {Coauthorship
  networks and patterns of scientific collaboration},}\ }\href@noop {}
  {\bibfield  {journal} {\bibinfo  {journal} {Proceedings of the national
  academy of sciences}\ }\textbf {\bibinfo {volume} {101}},\ \bibinfo {pages}
  {5200--5205} (\bibinfo {year} {2004})}\BibitemShut {NoStop}%
\bibitem [{\citenamefont {Grossman}(2002)}]{grossman2002evolution}%
  \BibitemOpen
  \bibfield  {author} {\bibinfo {author} {\bibfnamefont {Jerrold~W}\
  \bibnamefont {Grossman}},\ }\bibfield  {title} {\enquote {\bibinfo {title}
  {The evolution of the mathematical research collaboration graph},}\
  }\href@noop {} {\bibfield  {journal} {\bibinfo  {journal} {Congressus
  Numerantium}\ ,\ \bibinfo {pages} {201--212}} (\bibinfo {year}
  {2002})}\BibitemShut {NoStop}%
\bibitem [{\citenamefont {Moody}(2004)}]{Moody2004}%
  \BibitemOpen
  \bibfield  {author} {\bibinfo {author} {\bibfnamefont {James}\ \bibnamefont
  {Moody}},\ }\bibfield  {title} {\enquote {\bibinfo {title} {{The Structure of
  a Social Science Collaboration Network: Disciplinary Cohesion from 1963 to
  1999}},}\ }\href {\doibase 10.1177/000312240406900204} {\bibfield  {journal}
  {\bibinfo  {journal} {American Sociological Review}\ }\textbf {\bibinfo
  {volume} {69}},\ \bibinfo {pages} {213--238} (\bibinfo {year}
  {2004})}\BibitemShut {NoStop}%
\bibitem [{\citenamefont {Powell}\ \emph {et~al.}(2005)\citenamefont {Powell},
  \citenamefont {White}, \citenamefont {Koput},\ and\ \citenamefont
  {Owen-Smith}}]{powell2005network}%
  \BibitemOpen
  \bibfield  {author} {\bibinfo {author} {\bibfnamefont {Walter~W}\
  \bibnamefont {Powell}}, \bibinfo {author} {\bibfnamefont {Douglas~R}\
  \bibnamefont {White}}, \bibinfo {author} {\bibfnamefont {Kenneth~W}\
  \bibnamefont {Koput}}, \ and\ \bibinfo {author} {\bibfnamefont {Jason}\
  \bibnamefont {Owen-Smith}},\ }\bibfield  {title} {\enquote {\bibinfo {title}
  {Network dynamics and field evolution: The growth of interorganizational
  collaboration in the life sciences1},}\ }\href@noop {} {\bibfield  {journal}
  {\bibinfo  {journal} {American journal of sociology}\ }\textbf {\bibinfo
  {volume} {110}},\ \bibinfo {pages} {1132--1205} (\bibinfo {year}
  {2005})}\BibitemShut {NoStop}%
\bibitem [{\citenamefont {Goyal}\ \emph {et~al.}(2006)\citenamefont {Goyal},
  \citenamefont {{Van Der Leij}},\ and\ \citenamefont
  {Moraga-Gonzalez}}]{Goyal2006}%
  \BibitemOpen
  \bibfield  {author} {\bibinfo {author} {\bibfnamefont {Sanjeev}\ \bibnamefont
  {Goyal}}, \bibinfo {author} {\bibfnamefont {Marco~J}\ \bibnamefont {{Van Der
  Leij}}}, \ and\ \bibinfo {author} {\bibfnamefont {Jose~Luis}\ \bibnamefont
  {Moraga-Gonzalez}},\ }\bibfield  {title} {\enquote {\bibinfo {title}
  {{Economics: An Emerging Small World}},}\ }\href {\doibase 10.1086/500990}
  {\bibfield  {journal} {\bibinfo  {journal} {Journal of Political Economy}\
  }\textbf {\bibinfo {volume} {114}},\ \bibinfo {pages} {403--412} (\bibinfo
  {year} {2006})}\BibitemShut {NoStop}%
\bibitem [{\citenamefont {Lee}\ \emph {et~al.}(2008)\citenamefont {Lee},
  \citenamefont {Kahng},\ and\ \citenamefont {Goh}}]{lee2008evolution}%
  \BibitemOpen
  \bibfield  {author} {\bibinfo {author} {\bibfnamefont {Deokjae}\ \bibnamefont
  {Lee}}, \bibinfo {author} {\bibfnamefont {B}~\bibnamefont {Kahng}}, \ and\
  \bibinfo {author} {\bibfnamefont {Kwang-Il}\ \bibnamefont {Goh}},\ }\bibfield
   {title} {\enquote {\bibinfo {title} {Evolution of the coauthorship
  network},}\ }\href@noop {} {\bibfield  {journal} {\bibinfo  {journal}
  {Journal of the Korean Physical Society}\ }\textbf {\bibinfo {volume} {52}}
  (\bibinfo {year} {2008})}\BibitemShut {NoStop}%
\bibitem [{\citenamefont {Heidler}(2011)}]{heidler2011cognitive}%
  \BibitemOpen
  \bibfield  {author} {\bibinfo {author} {\bibfnamefont {Richard}\ \bibnamefont
  {Heidler}},\ }\bibfield  {title} {\enquote {\bibinfo {title} {Cognitive and
  social structure of the elite collaboration network of astrophysics: a case
  study on shifting network structures},}\ }\href@noop {} {\bibfield  {journal}
  {\bibinfo  {journal} {Minerva}\ }\textbf {\bibinfo {volume} {49}},\ \bibinfo
  {pages} {461--488} (\bibinfo {year} {2011})}\BibitemShut {NoStop}%
\bibitem [{\citenamefont {Dewey}(1910)}]{dewey1910science}%
  \BibitemOpen
  \bibfield  {author} {\bibinfo {author} {\bibfnamefont {John}\ \bibnamefont
  {Dewey}},\ }\bibfield  {title} {\enquote {\bibinfo {title} {Science as
  subject-matter and as method},}\ }\href@noop {} {\bibfield  {journal}
  {\bibinfo  {journal} {Science}\ ,\ \bibinfo {pages} {121--127}} (\bibinfo
  {year} {1910})}\BibitemShut {NoStop}%
\bibitem [{\citenamefont {Arons}\ and\ \citenamefont
  {Karplus}(1976)}]{Arons1976b}%
  \BibitemOpen
  \bibfield  {author} {\bibinfo {author} {\bibfnamefont {Arnold~B}\
  \bibnamefont {Arons}}\ and\ \bibinfo {author} {\bibfnamefont {R}~\bibnamefont
  {Karplus}},\ }\bibfield  {title} {\enquote {\bibinfo {title} {{Implications
  of Accumulating Data on Levels of Intellectual Development}},}\ }\href@noop
  {} {\bibfield  {journal} {\bibinfo  {journal} {American Journal of Physics}\
  }\textbf {\bibinfo {volume} {44}} (\bibinfo {year} {1976})}\BibitemShut
  {NoStop}%
\bibitem [{\citenamefont {Arons}(1976)}]{Arons1976}%
  \BibitemOpen
  \bibfield  {author} {\bibinfo {author} {\bibfnamefont {A~B}\ \bibnamefont
  {Arons}},\ }\bibfield  {title} {\enquote {\bibinfo {title} {{Cultivating the
  capacity for formal reasoning}},}\ }\href@noop {} {\bibfield  {journal}
  {\bibinfo  {journal} {American Journal of Physics}\ }\textbf {\bibinfo
  {volume} {44}},\ \bibinfo {pages} {834} (\bibinfo {year} {1976})}\BibitemShut
  {NoStop}%
\bibitem [{\citenamefont {Barab{\^a}si}\ \emph {et~al.}(2002)\citenamefont
  {Barab{\^a}si}, \citenamefont {Jeong}, \citenamefont {N{\'e}da},
  \citenamefont {Ravasz}, \citenamefont {Schubert},\ and\ \citenamefont
  {Vicsek}}]{barabasi2002evolution}%
  \BibitemOpen
  \bibfield  {author} {\bibinfo {author} {\bibfnamefont {Albert-Laszlo}\
  \bibnamefont {Barab{\^a}si}}, \bibinfo {author} {\bibfnamefont {Hawoong}\
  \bibnamefont {Jeong}}, \bibinfo {author} {\bibfnamefont {Zoltan}\
  \bibnamefont {N{\'e}da}}, \bibinfo {author} {\bibfnamefont {Erzsebet}\
  \bibnamefont {Ravasz}}, \bibinfo {author} {\bibfnamefont {Andras}\
  \bibnamefont {Schubert}}, \ and\ \bibinfo {author} {\bibfnamefont {Tamas}\
  \bibnamefont {Vicsek}},\ }\bibfield  {title} {\enquote {\bibinfo {title}
  {Evolution of the social network of scientific collaborations},}\ }\href@noop
  {} {\bibfield  {journal} {\bibinfo  {journal} {Physica A: Statistical
  mechanics and its applications}\ }\textbf {\bibinfo {volume} {311}},\
  \bibinfo {pages} {590--614} (\bibinfo {year} {2002})}\BibitemShut {NoStop}%
\bibitem [{\citenamefont {Acedo}\ \emph {et~al.}(2006)\citenamefont {Acedo},
  \citenamefont {Barroso}, \citenamefont {Casanueva},\ and\ \citenamefont
  {Gal\'{a}n}}]{Acedo2006a}%
  \BibitemOpen
  \bibfield  {author} {\bibinfo {author} {\bibfnamefont {F~J}\ \bibnamefont
  {Acedo}}, \bibinfo {author} {\bibfnamefont {C}~\bibnamefont {Barroso}},
  \bibinfo {author} {\bibfnamefont {C}~\bibnamefont {Casanueva}}, \ and\
  \bibinfo {author} {\bibfnamefont {J~L}\ \bibnamefont {Gal\'{a}n}},\
  }\bibfield  {title} {\enquote {\bibinfo {title} {{Co-authorship in management
  and organizational studies: an empirical and network analysis}},}\ }\href
  {http://dx.doi.org/10.1111/j.1467-6486.2006.00625.x} {\bibfield  {journal}
  {\bibinfo  {journal} {Journal of Management Studies}\ }\textbf {\bibinfo
  {volume} {43}},\ \bibinfo {pages} {957--983} (\bibinfo {year}
  {2006})}\BibitemShut {NoStop}%
\bibitem [{\citenamefont {Banerjee}\ \emph {et~al.}(2013)\citenamefont
  {Banerjee}, \citenamefont {Chandrasekhar}, \citenamefont {Duflo},\ and\
  \citenamefont {Jackson}}]{banerjee2013diffusion}%
  \BibitemOpen
  \bibfield  {author} {\bibinfo {author} {\bibfnamefont {Abhijit}\ \bibnamefont
  {Banerjee}}, \bibinfo {author} {\bibfnamefont {Arun~G}\ \bibnamefont
  {Chandrasekhar}}, \bibinfo {author} {\bibfnamefont {Esther}\ \bibnamefont
  {Duflo}}, \ and\ \bibinfo {author} {\bibfnamefont {Matthew~O}\ \bibnamefont
  {Jackson}},\ }\bibfield  {title} {\enquote {\bibinfo {title} {The diffusion
  of microfinance},}\ }\href@noop {} {\bibfield  {journal} {\bibinfo  {journal}
  {Science}\ }\textbf {\bibinfo {volume} {341}},\ \bibinfo {pages} {1236498}
  (\bibinfo {year} {2013})}\BibitemShut {NoStop}%
\bibitem [{Note1()}]{Note1}%
  \BibitemOpen
  \bibinfo {note} {Biomedical, astrophysics, condensed matter physics, and
  high-energy theory: \cite {Newman2001PNAS} Sociology: \cite {Moody2004}
  Management: \cite {Acedo2006a}}\BibitemShut {NoStop}%
\bibitem [{\citenamefont {Cunningham}\ and\ \citenamefont
  {Dillon}(1997)}]{cunningham1997authorship}%
  \BibitemOpen
  \bibfield  {author} {\bibinfo {author} {\bibfnamefont {Sally~Jo}\
  \bibnamefont {Cunningham}}\ and\ \bibinfo {author} {\bibfnamefont {Stuart~M}\
  \bibnamefont {Dillon}},\ }\bibfield  {title} {\enquote {\bibinfo {title}
  {Authorship patterns in information systems},}\ }\href@noop {} {\bibfield
  {journal} {\bibinfo  {journal} {Scientometrics}\ }\textbf {\bibinfo {volume}
  {39}},\ \bibinfo {pages} {19--27} (\bibinfo {year} {1997})}\BibitemShut
  {NoStop}%
\bibitem [{Note2()}]{Note2}%
  \BibitemOpen
  \bibinfo {note} {The 95\% confidence interval for the point estimate of the
  log odds ratio is $ln(y)\pm 1.96*ste(ln(y))$, where $ste(ln(y))=\protect
  \sqrt {\protect \frac {1}{a_{1}}+\protect \frac {1}{b_{1}}+\protect \frac
  {1}{a_{2}}+\protect \frac {1}{b_{2}}}$.}\BibitemShut {Stop}%
\bibitem [{\citenamefont {Barabasi}\ \emph {et~al.}(2002)\citenamefont
  {Barabasi}, \citenamefont {Jeong}, \citenamefont {Neda}, \citenamefont
  {Ravasz}, \citenamefont {Schubert},\ and\ \citenamefont
  {Vicsek}}]{BARABASI2002}%
  \BibitemOpen
  \bibfield  {author} {\bibinfo {author} {\bibfnamefont {A-L}\ \bibnamefont
  {Barabasi}}, \bibinfo {author} {\bibfnamefont {H}~\bibnamefont {Jeong}},
  \bibinfo {author} {\bibfnamefont {Z}~\bibnamefont {Neda}}, \bibinfo {author}
  {\bibfnamefont {E}~\bibnamefont {Ravasz}}, \bibinfo {author} {\bibfnamefont
  {A}~\bibnamefont {Schubert}}, \ and\ \bibinfo {author} {\bibfnamefont
  {T}~\bibnamefont {Vicsek}},\ }\bibfield  {title} {\enquote {\bibinfo {title}
  {{Evolution of the social network of scientific collaborations}},}\ }\href
  {\doibase 10.1016/S0378-4371(02)00736-7} {\bibfield  {journal} {\bibinfo
  {journal} {Physica A: Statistical Mechanics and its Applications}\ }\textbf
  {\bibinfo {volume} {311}},\ \bibinfo {pages} {590--614} (\bibinfo {year}
  {2002})}\BibitemShut {NoStop}%
\bibitem [{\citenamefont {{AIP Statistical Research
  Center}}(2015)}]{AIP2015FirstYear}%
  \BibitemOpen
  \bibfield  {author} {\bibinfo {author} {\bibnamefont {{AIP Statistical
  Research Center}}},\ }\href@noop {} {\enquote {\bibinfo {title} {{First-Year
  Graduate Physics Students}},}\ } (\bibinfo {year} {2015})\BibitemShut
  {NoStop}%
\bibitem [{\citenamefont {Meltzer}\ and\ \citenamefont
  {Thornton}(2012)}]{Meltzer2012}%
  \BibitemOpen
  \bibfield  {author} {\bibinfo {author} {\bibfnamefont {David~E.}\
  \bibnamefont {Meltzer}}\ and\ \bibinfo {author} {\bibfnamefont {Ronald~K.}\
  \bibnamefont {Thornton}},\ }\bibfield  {title} {\enquote {\bibinfo {title}
  {{Resource Letter ALIP--1: Active-Learning Instruction in Physics}},}\ }\href
  {\doibase 10.1119/1.3678299} {\bibfield  {journal} {\bibinfo  {journal}
  {American Journal of Physics}\ }\textbf {\bibinfo {volume} {80}},\ \bibinfo
  {pages} {478} (\bibinfo {year} {2012})}\BibitemShut {NoStop}%
\bibitem [{\citenamefont {Madsen}\ \emph {et~al.}(2017)\citenamefont {Madsen},
  \citenamefont {McKagan},\ and\ \citenamefont {Sayre}}]{Madsen2017RBAI}%
  \BibitemOpen
  \bibfield  {author} {\bibinfo {author} {\bibfnamefont {Adrian}\ \bibnamefont
  {Madsen}}, \bibinfo {author} {\bibfnamefont {Sam}\ \bibnamefont {McKagan}}, \
  and\ \bibinfo {author} {\bibfnamefont {Eleanor~C}\ \bibnamefont {Sayre}},\
  }\bibfield  {title} {\enquote {\bibinfo {title} {{Resource Letter: RBAI-1:
  Research-based Assessment Instruments in Physics and Astronomy}},}\ }\href
  {http://arxiv.org/abs/1605.02703} {\bibfield  {journal} {\bibinfo  {journal}
  {American Journal of Physics}\ } (\bibinfo {year} {2017})},\ \Eprint
  {http://arxiv.org/abs/1605.02703} {arXiv:1605.02703} \BibitemShut {NoStop}%
\end{thebibliography}
%

\end{document}